\documentclass[useAMS,usenatbib,psfig]{mn2e}
\usepackage{graphicx}
\usepackage{natbib}
\newcommand{\hii}{{H{\scriptsize II} }}

\newcommand{\kms}{\mbox{kms$^{-1}$}}

\newcommand{\nhone}{\mbox{NH$_3$(1,1)}}

\newcommand{\nhfour}{\mbox{NH$_3$(4,4)}}

\title[Extended free-free emission towards MSF regions]{Too large and
  overlooked? Extended free-free emission towards massive star
  formation regions}
\author[S.N.Longmore et al.]{S. N. Longmore$^{1,2,3}$\thanks{E-mail:
slongmore@cfa.harvard.edu}, M. G. Burton$^2$, E. Keto$^{1}$, S. Kurtz$^{4}$ \& A. J. Walsh$^5$\\
$^{1}$Harvard-Smithsonian Center for Astrophysics, 60 Garden Street, Cambridge, MA 02138, USA\\
$^{2}$School of Physics, University of New South Wales, 2052, Sydney, Australia\\
$^{3}$Australia Telescope National Facility, CSIRO, Epping 1710, Sydney, Australia\\
$^{4}$CRyA, Universidad Nacional Aut\'onoma de M\'exico, Apartado Postal 3-72, 58089, Morelia, Michoac\'{a}n, M\'{e}xico\\
$^{5}$Centre for Astronomy, James Cook University, Townsville, QLD 4811 Australia
}

\begin{document}

\date{in prep.}

\pagerange{\pageref{firstpage}--\pageref{lastpage}} \pubyear{2009}

\maketitle

\begin{abstract}

We present Australia Telescope Compact Array observations towards 6
massive star formation regions which, from their strong 24\,GHz
continuum emission but no compact 8\,GHz continuum emission, appeared
good candidates for hyper-compact \hii~regions. However, the
properties of the ionised gas derived from the 19 to 93\,GHz continuum
emission and H70$\alpha$ $+$ H57$\alpha$ radio recombination line data
show the majority of these sources are, in fact, regions of
spatially-extended, optically-thin free-free emission. These extended
sources were missed in the previous 8\,GHz observations due to a
combination of spatial-filtering, poor surface brightness sensitivity
and primary beam attenuation.

We consider the implications that a significant number of these
extended \hii~regions may have been missed by previous surveys of
massive star formation regions. If the original sample of 21 sources
is representative of the population as a whole, the fact that 6
contain previously undetected extended free-free emission suggests a
large number of regions have been mis-classified. Rather than being
very young objects prior to UC\hii~region formation, they are, in
fact, associated with extended \hii~regions and thus significantly
older. In addition, inadvertently ignoring a potentially substantial
flux contribution (up to $\sim$0.5Jy) from free-free emission has
implications for dust masses derived from sub-mm flux densities. The
large spatial scales probed by single-dish telescopes, which do not
suffer from spatial filtering, are particularly susceptible and dust
masses may be overestimated by up to a factor of $\sim$2.

\end{abstract}


\section{Introduction}
A definition often used to describe a star as `massive' requires it to
have sufficient UV radiation to ionise the surrounding
environment. The ionised material can then be observed as free-free
continuum emission at radio wavelengths. When a massive star is
forming, the emergence and subsequent expansion of detectable
free-free emission from the natal cocoon, provides an indicator of the
stars evolutionary state. As the ionising radiation overcomes
quenching due to the large mass of infalling material, the continuum
emission emerges. At first the ionised gas is extremely dense, but
once the ionised gas pressure is sufficient to overcome the
gravitational potential, the \hii~region expands into ultra compact
(UC), compact and finally diffuse \hii~regions. The spectral energy
distribution (SED) at cm wavelengths falls into two distinct regimes
with increasing frequency. As the free-free emission turns from
optically thick to thin, the spectral index (S$_\nu \propto
\nu^\alpha$, where S$_\nu$ is the flux density at frequency $\nu$)
changes from $\alpha \sim$ 2 to -0.1. The frequency at which the
turnover occurs scales almost linearly with the electron density
\citep{kurtz2005} i.e., denser and presumably younger objects will
exhibit the turn over at higher frequencies. At shorter wavelengths
the SED quickly becomes dominated by thermal emission from dust in the
surrounding natal cocoon, which has a spectral index, $\alpha \sim 3 -
4$.

To identify sources at the earliest stages of massive star formation
we used the work of \citet[][hereafter WBHR98]{walsh1998} to select
objects in the relatively unexplored southern sky with: (i) SED's
peaking in the far-infrared, (ii) sufficient luminosity to expect
stars massive enough to ionise the surrounding material, (iii) no
detectable cm free-free emission and (iv) class II methanol maser
emission \citep[][hereafter L07]{L07A}. Class II methanol masers are
generally found offset from cm continuum emission but always
coincident with sub-mm cores \citep[e.g.][]{walsh2003} which have
sufficient luminosity to contain massive stars (WBHR98). These highly
luminous cores with no detected cm emission are interpreted as massive
star forming cores prior to the onset of UC\hii~formation. Subsequent
multi-wavelength studies have since derived the large-scale properties
(mass, chemical composition, temperature etc.) of these star forming
cores using the SEST, Mopra and ATCA telescopes \citep{hill2005,
  purcell2006, purcell2009, L07B, longmore2008} and performed a census
of the (proto-) stellar populations at infrared wavelengths
\citep{longmore2006, L07_I2}.

In L07 we selected 21 regions for further study, all of which
contained methanol maser emission detected by WBHR98. Six of these
regions were reported as containing continuum emission by the
simultaneously-observed 8\,GHz observations of WHBR98, while the
remaining 15 were reported as devoid of 8\,GHz continuum emission by
their 8\,GHz observations. Using this selection criteria, we hoped to
identify 15 young regions prior to UCHII formation and 6 older
regions, post-UCHII formation. However, contrary to expectation,
twelve rather than six of these regions were detected in the 24 GHz
continuum by L07: the six detected in the 8\,GHz continuum by WBHR98
and six reported as devoid of 8\,GHz continuum by WBHR98.

In this work we are seeking to find the nature of the sources detected
at 24\,GHz by L07 but not at 8\,GHz by WBHR98. If these cores are
indeed at an earlier evolutionary stage as originally thought, this
emission could be free-free emission in the optically-thick regime
between 8 and 24\,GHz. In this case their 24\,GHz flux should be
$\sim$10 times brighter than at 8\,GHz and may explain the 8\,GHz
non-detections. However, both L07 and WBHR98 observations were made
using an interferometer with significantly different array
configurations and so probed the source structure at different spatial
frequencies and with different surface brightness sensitivities. The
other possibility discussed in L07 is that these sources could instead
be much older, spatially extended, optically thin \hii~regions which
were resolved out/not-detected due to the extended array configuration
used by WBHR98. \citet{kurtz1999} found just such extended continuum
emission surrounding 12 out of 15 UC\hii~regions they observed, with
important consequences for interpreting the nature of these regions.
If such a result was found for many of the L07 sources, and these were
representative of the WBHR98 sample as a whole, this would have
ramifications regarding their currently assumed age and evolutionary
state. Not only would this imply the sources are significantly older
than originally thought, but it may also affect attempts to derive
dust masses at shorter wavelengths ignoring the contribution from the
extended free-free emission. In this paper we aim to uncover the
nature of these sources through continuum and radio recombination line
observations from 19 to 93GHz.

\section{Observations and Data Reduction}
\label{sec:obs_red}

\begin{table}
  \caption{Pointing centre from the methanol maser position, V$_{\rm
      LSR}$ \& IRAS name (from WBHR98) and the adopted distance
    \citep[from][]{purcell2006} of the observed regions.  }
  \begin{tabular}{|c|c|c|c|c|c|} \hline \hline

Region       & IRAS         & RA         & DEC       & V$_{\rm LSR}$ & D \\
             & name         & (J2000)    & (J2000)   & (km/s)        &(kpc) \\ \hline
G316.81-0.06 & 14416$-$5937 & 14:45:26.9 & -59:49:16 & -38.7         & 2.7 \\
G331.28-0.19 & 16076$-$5134 & 16:11:26.9 & -51:41:57 & -88.1         & 5.4 \\
G12.68-0.18  & 18117$-$1753 & 18:13:54.7 & -18:01:41 & 56.5          & 4.9 \\
G19.47+0.17  & 18232$-$1154 & 18:25:54.7 & -11:52:34 & 19.7          & 1.9 \\
G24.79+0.08  & 18335$-$0711 & 18:36:12.3 & -07:12:11 & 110.5         & 7.7 \\
G24.85+0.09  & 18335$-$0711 & 18:36:18.4 & -07:08:52 & 108.9         & 6.3 \\\hline

  \end{tabular}

\label{tab:regions}
\end{table}

The sources were selected from L07 as those with 24\,GHz continuum
detections but no corresponding 8\,GHz detections reported in
WBHR98. Table~\ref{tab:regions} lists the pointing centres from the
methanol maser position, V$_{\rm LSR}$ and kinematic distances
\citep[from][]{purcell2006} for each of the regions.

\subsection{ATCA continuum and radio recombination line (RRL) data}
The Australia Telescope Compact Array (ATCA) observations were
undertaken between 2006 and 2008 as part of ATCA projects C1287 and
C1751. The observing setups (array and correlator settings,
approximate sensitivities and beam sizes, continuum frequencies and
RRL transitions) are listed in Table~\ref{tab:obs_setup}. In order to
mitigate problems constructing SEDs of extended sources with different
beam sizes, three array configurations were used (H75, H168 and H214,
with maximum baselines of 75m, 168m and 214m, respectively) in an
attempt to match the beams at the different observing
frequencies. These configurations contain both north-south plus
east-west baselines to allow for snapshot imaging. Each source was
observed in multiple 15 minute cuts, with a close phase calibrator
observed for 3 minutes before and after each cut. PKS~1253-055 and
PKS~1921-293 were used as the bandpass calibrators. The ATCA standard
PKS~1934-634 was used as the primary calibrator for observations below
30\,GHz and Uranus was used at higher frequencies. The data were
reduced in the standard way using the MIRIAD \citep{sault1995}
package. Bad visibilities were flagged and the gains/bandpass
solutions from the appropriate calibrator were applied to the
visibilities. A zeroeth order polynomial was fit to the line-free
channels and subtracted from the visibilities to separate the line and
continuum emission. These separate visibilities were then Fourier
transformed to form image cubes. Due to variation in system
temperatures between the antennae, individual visibilities were
weighted by their system temperature. The images were cleaned and
convolved with a Gaussian synthesised beam to produce a final restored
image.

Continuum fluxes were extracted using \emph{imfit} on the restored
images. Extended sources were fit with a 2D Gaussian profile to
calculate the spatial extent, integrated and peak fluxes
densities. Unresolved emission was fit as a point source. The
positions, peak flux and deconvolved sizes of the continuum emission
from these fits at each frequency and array configuration are shown in
Table~\ref{tab:cont_properties}. Where detected, RRL spectra were
extracted for each source at the peak of the continuum emission,
baseline subtracted and fit with a Gaussian profile using the
\emph{gauss} method in
CLASS\footnote{http://www.iram.fr/GILDAS}. Table~\ref{tab:rrl_properties}
lists the velocity-integrated flux, V$_{\rm LSR}$, FWHM, peak flux and
their corresponding uncertainties for each of the detected transitions
and array configurations.

\begin{table*}
    \caption{Overview of the observing setup. For the H70$\alpha$,
      H57$\alpha$ and H41$\alpha$ observations, the observing
      frequencies were centred on the respective RRL transitions and
      continuum emission was extracted from line-free channels. The
      93.5\,GHz observations were solely high spatial resolution
      continuum observations and no RRL was observed in this
      setup. Columns three and four give the bandwidth (BW) and number
      of channels for each of the correlator settings used. $\Delta$V
      shows the corresponding velocity resolution per channel for each
      IF in $\kms$. Column six lists the array configuration for each
      observation. The H75, H168 and H214 array configurations have
      baselines ranging from 31 - 89\,m, 61 - 192\,m and 82 - 247\,m,
      respectively. Columns 7 and 8 show the theoretical spectral line
      RMS per channel and theoretical continuum brightness sensitivity
      (based on a one hour observation in average weather conditions
      using the ATCA sensitivity calculator --
      http://www.atnf.csiro.au). The final column gives a
      characteristic synthesised beam size (for a source at $\delta =
      -30^\circ$).}
  \begin{center}
    \begin{tabular}{|c|c|c|c|c|c|c|c|c|c|c|c|c|}\hline\hline      
 Trans.        & Rest Freq.& BW    &  Num. & $\Delta$V &  Array & Line RMS        & Cont. RMS  &  Beam \\ 
               & (GHz)     & (MHz) &  Chan & ($\kms$)  & Config & (mJy/beam/chan) & (mJy/beam) & ($\arcsec$)   \\ \hline
 H70$\alpha$   & 18.769    &  16   &  512  & 0.50     &  H168  & 7                & 0.3        &  14    \\ 
 H57$\alpha$   & 34.596    &  32   &  256  & 1.08     &  H75   & 6                & 0.4        &  18   \\
               &           &       &       &           &  H168  & 6               & 0.4        &  8   \\
 H41$\alpha$   & 92.034    &  32   &  256  & 0.41     &  H75   & 56               & 4          &  7   \\ 
 cont.         & 93.5      &  128  &  32   & 26        &  H214  & -               & 2          &  3 \\\hline   
    \end{tabular}

    \label{tab:obs_setup}
  \end{center}
  \vspace{-2mm}
\end{table*}

\begin{table*}
  \begin{center}
    \caption{Continuum detections towards each of the regions. The
      frequency and array configurations are given in columns two and
      three. The listed positions, fluxes and angular sizes are taken
      from the fits to the images, as outlined in
      $\S$~\ref{sec:obs_red}. For extended sources, the angular sizes
      have been deconvolved with the synthesised beam. The errors
      shown in column seven are from the fit to the data and do not
      take into account the absolute error in the flux scale which is
      assumed to be 20\% and 30\% for the observations $<90$\,GHz and
      $>90$\,GHz, respectively. $^a$ the integrated flux for extended
      sources is shown in column seven in parentheses. $^b$ for
      integrated fluxes (in parentheses) the units are mJy rather than
      mJy/beam. $^c$ G316.81-0.06 Core 1 is considerably outside the
      primary beam of the 92\,GHz H75 observations making this flux
      measurement highly uncertain.}

    \begin{tabular}{|cccccccccc|} \hline \hline

Source               & Freq   & Array &RA         & DEC         &  Beam        &  Peak Flux$^a$ & $\theta_{RA}$ & $\theta_{DEC}$ & PA        \\
                     & (GHz)  &       &(J2000)    & (J2000)     &  ($\arcsec$) & (mJy/beam$^b$)          & ($\arcsec$)   & ($\arcsec$)    & ($^\circ$)\\ \hline

G316.81-0.06 Core 1  & 92.0 & H75   & 14:45:29.26 & -59:49:09.1 & 2.6$\times$1.4   & (99)43$\pm$19$^c$& 10 & 5  & $-$5 \\
G316.81-0.06 Core 2a & 18.7 & H168  & 14:45:25.59 & -59:49:12.7 & 13.1$\times$9.7  & (456)284$\pm$77  & 11 & 6  & -3 \\
                     & 24.0 & H168  & 14:45:25.68 & -59:49:15.2 & 10.2$\times$9.3  &  435$\pm$64      & -  & -  & - \\
                     & 34.6 & H75   & 14:45:25.45 & -59:49:10.6 & 16.3$\times$13.0 & (766)422$\pm$337 & 14 & 11 & 16 \\
                     & 34.6 & H168  & 14:45:25.91 & -59:49:15.4 & 7.8$\times$5.9   & (168)72$\pm$27   & 10 & 5  & 5 \\
                     & 92.0 & H75   & 14:45:26.26 & -59:49:16.9 & 7.3$\times$6.0   & (135)57$\pm$16   & 10 & 5  & -29\\
G316.81-0.06 Core 2b & 18.7 & H168  & 14:45:25.48 & -59:49:27.4 & 13.1$\times$9.7  & (598)334$\pm$77  & 11 & 8  & -4 \\
                     & 24.0 & H168  & 14:45:26.04 & -59:49:24.4 & 10.2$\times$9.3  & 496$\pm$65       & -  & -  & - \\
                     & 34.6 & H75   & 14:45:25.68 & -59:49:25.7 & 16.3$\times$13.0 & (829)490$\pm$441 & 13 & 10 & 46 \\
                     & 34.6 & H168  & 14:45:25.75 & -59:49:27.8 & 7.8$\times$5.9   & (177)67$\pm$14   & 13 & 3  & 1 \\
                     & 92.0 & H75   & 14:45:25.70 & -59:49:26.7 & 7.3$\times$6.0   & (36)30$\pm$23    & 9  & 6  & 9 \\
G316.81-0.06 Core 3a & 18.7 & H168  & 14:45:22.40 & -59:49:35.8 & 13.1$\times$9.7  & (490)451$\pm$99  & -  & -  & -   \\
                     & 24.0 & H168  & 14:45:21.98 & -59:49:41.5 & 10.2$\times$9.3  & (1199)432$\pm$59 & 14 & 12  & -17 \\
                     & 34.6 & H75   & 14:45:22.02 & -59:49:35.0 & 16.3$\times$13.0 & (480)216$\pm$371 & 27 & 6  & 80 \\
                     & 34.6 & H168  & 14:45:22.61 & -59:49:37.1 & 7.8$\times$5.9   & (459)121$\pm$77  & 12 & 11 & -61 \\ 
G316.81-0.06 Core 3b & 18.7 & H168  & 14:45:20.86 & -59:49:27.4 & 13.1$\times$9.7  & (901)297$\pm$42  & 17 & 15 & -20 \\
                     & 24.0 & H168  & 14:45:20.58 & -59:49:28.4 & 10.2$\times$9.3  & (644)405$\pm$75  & 11 & 6  & 2.1 \\
                     & 34.6 & H75   & 14:45:21.57 & -59:49:26.7 & 16.3$\times$13.0 & (248)122$\pm$279 & 22 & 7  & -35 \\
                     & 34.6 & H168  & 14:45:21.40 & -59:49:29.7 & 7.8$\times$5.9   & (32)22$\pm$12    & 6  & 1  & -7 \\

G331.28-0.19 Core 1  & 18.7 & H168  & 16:11:27.36 & -51:41:59.7 & 13.2$\times$9.5  & (296)86$\pm$3    & 25 & 12 & $-$84\\ 
                     & 24.0 & H168  & 16:11:27.6 & -51:41:59.68 & 13.07$\times$8.9 & (280)80$\pm$10   & 23 & 17 & $-$79 \\
                     & 34.6 & H75   & 16:11:27.30 & -51:41:59.4 & 18.6$\times$11.9 & (299)146$\pm$4   & 20 & 11 & $-$67 \\
                     & 34.6 & H168  & 16:11:27.47 & -51:41:59.2 &  9.2$\times$5.6  & (265)44$\pm$2    & 25 & 10 & $-$90 \\  
                     & 92.0 & H75   & 16:11:26.63 & -51:41:56.6 & 6.5$\times$4.9   & 73$\pm$23        & -  & -  & -     \\ 
                     & 93.5 & H214  & 16:11:26.57 & -51:41:57.2 & 2.6$\times$1.5   & (29)11$\pm$1     & 3  & 2  & -28   \\ 

G12.68-0.18 Core 3   & 18.7 & H168  & 18:13:55.40 & -18:01:38.4 & 14.0$\times$8.7  & (274)74$\pm$3    & 8  & 17 & 18    \\ 
                     & 24.0 & H168  & 18:13:55.30 & -18:01:36.2 & 12.0$\times$8.2  & 140$\pm$10       & 13  & 9  & 27   \\
                     & 34.6 & H75   & 18:13:55.30 & -18:01:37.1 & 20.9$\times$12.3 & 42$\pm$3         & -  & -  & -     \\
                     & 34.6 & H168  & 18:13:55.29 & -18:01:35.8 & 8.1$\times$5.2   & (79)27$\pm$3     & 14 & 6  & 50    \\    
                     & 92.0 & H75   & 18:13:54.73 & -18:01:38.7 & 6.8$\times$4.8   & 37$\pm$9         & -  & -  & -     \\
G12.68-0.18 Core 4   & 92.0 & H75   & 18:13:54.56 & -18:01:44.9 & 6.8$\times$4.8   & 32$\pm$10        & -  & -  & -     \\
                     & 93.5 & H214  & 18:13:54.74 & -18:01:46.3 & 3.0$\times$1.4   & 26$\pm$3         & -  & -  & -     \\

G19.47+0.17 Core 1   & 92.0 & H75   & 18:25:54.75 & -11:52:34.2 & 7.0$\times$5.0   & (92)58$\pm$9     & 5  & 4  & 71    \\
                     & 93.5 & H214  & 18:25:54.72 & -11:52:34.0 & 3.1$\times$1.5   & 33$\pm$2         & -  & -  & -     \\
G19.47+0.17 Core 2   & 18.7 & H168  & 18:25:54.34 & -11:52:22.3 & 14.1$\times$9.1  & 34$\pm$3         & -  & -  & -     \\
                     & 34.6 & H75   & 18:25:54.39 & -11:52:22.8 & 19.4$\times$12.4 & 31$\pm$3         & -  & -  & -     \\
                     & 34.6 & H168  & 18:25:54.30 & -11:52:21.7 & 9.7$\times$5.2   & 26$\pm$3         & -  & -  & -     \\
                     & 92.0 & H75   & 18:25:54.35 & -11:52:21.0 & 7.0$\times$5.0   & 26$\pm$13        & -  & -  & -     \\

G24.79+0.08 Core 1   & 18.7 & H168  & 18:36:12.58 & -7:12:12.0  & 14.7$\times$9.0  & 77$\pm$12        & -  & -  & -     \\
                     & 34.6 & H75   & 18:36:12.54 & -7:12:11.7  & 19.4$\times$12.2 & 118$\pm$8        & -  & -  & -     \\
                     & 34.6 & H168  & 18:36:12.52 & -7:12:12.0  & 10.2$\times$5.1  & 104$\pm$5        & -  & -  & -     \\
                     & 92.0 & H75   & 18:36:12.60 & -7:12:10.4  & 7.2$\times$4.5   & (244)171$\pm$24  & 4  & 1  & $-$8  \\
                     & 93.5 & H214  & 18:36:12.56 & -7:12:11.0  & 3.6$\times$1.3   & 133$\pm$6        & -  & -  & -     \\   
G24.79+0.08 Core 3   & 18.7 & H168  & 18:36:10.45 & -7:11:21.1  & 14.7$\times$9.0  & 154$\pm$11       & -  & -  & -     \\  
                     & 34.6 & H75   & 18:36:10.53 & -7:11:26.9  & 19.4$\times$12.2 & 65$\pm$11        & -  & -  & -     \\           
                     & 34.6 & H168  & 18:36:10.66 & -7:11:21.5  & 10.2$\times$5.1  & (104)25$\pm$6    & 15 & 9  & $-$17 \\  

G24.85+0.09 Core 1   & 18.7 & H168  & 18:36:18.51 & -7:08:54.0  & 15.0$\times$9.0  & 69$\pm$6         & -  & -  & -     \\   
                     & 34.6 & H75   & 18:36:18.57 & -7:08:54.7  & 19.0$\times$13.1 & (116)53$\pm$5    & 29 & 19 & 74    \\
                     & 34.6 & H168  & 18:36:18.21 & -7:08:51.9  & 10.4$\times$5.0  & (62)24$\pm$2     & 11 & 5  & $-$40 \\  \hline

    \end{tabular}
    \label{tab:cont_properties}
  \end{center}
\end{table*}

\begin{table*}
  \begin{center}
    \caption{Detected radio recombination line emission towards each
      region. The second and third columns give the transition and
      array configuration. Columns four through seven list the
      velocity-integrated line flux, central velocity, full-width at
      half maximum and peak intensity with associated errors from fits
      to the spectra (see $\S$~\ref{sec:obs_red}), extracted from the
      position of peak continuum flux in
      Table~\ref{tab:cont_properties}. Column 8 gives the line to
      continuum ratio for each transition.  $^a$ the reported flux is
      the velocity-integrated flux extracted from the peak
      spectra. The spectra are shown for each region separately in
      Figures~\ref{fig:g316_spectra}, \ref{fig:g331_spectra},
      \ref{fig:g12.68_spectra}, \ref{fig:g19.47_spectra},
      \ref{fig:g24.79_spectra} and \ref{fig:g24.85_spectra} and
      overlayed with the fit to profile.}

    \begin{tabular}{|cccccccc|} \hline \hline

Source             & Trans.      & Array & Flux$^a$        &  V$_{\rm LSR}$  & FWHM          &  Peak        & Line/cont. \\
                   &             &       & (mJy/beam.km/s) &  (km/s)         & (km/s)        &  (mJy/beam)  & ratio      \\ \hline
G316.81-0.06 Core 2& H57$\alpha$ & H75   & 7610$\pm$54     & -43.0$\pm$0.08  & 23.7$\pm$0.20 & 300          & 0.5 \\
                   & H57$\alpha$ & H168  & 1660$\pm$54     & -46.5$\pm$0.34  & 20.8$\pm$0.84 & 72           & 0.5 \\
                   & H70$\alpha$ & H168  & 2483$\pm$26     & -43.8$\pm$0.11  & 20.8$\pm$0.26 & 110          & 0.3 \\
G316.81-0.06 Core 3& H57$\alpha$ & H75   & 3695$\pm$62     & -43.2$\pm$0.20  & 24.1$\pm$0.49 & 140          & 0.5\\
                   & H57$\alpha$ & H168  & 1630$\pm$48     & -45.5$\pm$0.28  & 19.4$\pm$0.68 & 79           & 0.5\\
                   & H70$\alpha$ & H168  & 2737$\pm$32     & -42.3$\pm$0.14  & 24.8$\pm$0.35 & 100          & 0.2\\
G331.28-0.19 Core 1& H57$\alpha$ & H75   & 1363$\pm$19     & -76.6$\pm$0.12  & 17.9$\pm$0.28 & 71           & 0.5\\
                   & H70$\alpha$ & H168  & 429$\pm$20      & -78.6$\pm$0.48  & 21.8$\pm$1.2 & 18            & 0.3\\
G12.68-0.18 Core 3 & H57$\alpha$ & H75   & 450$\pm$23      &  57.5$\pm$0.85  & 33.8$\pm$2.0  & 13           & 0.4\\
                   & H57$\alpha$ & H168  & 273$\pm$22      &  55.8$\pm$0.98  & 22.7$\pm$2.0  & 11           & 0.5\\
                   & H70$\alpha$ & H168  & 780$\pm$64      &  54.7$\pm$1.3  & 34.3$\pm$3.6  & 21            & 0.3 \\
G19.47+0.17 Core 2 & H57$\alpha$ & H75   & 429$\pm$53      &  16.4$\pm$1.2  & 20.9$\pm$3.2  & 19            & 0.6\\
                   & H70$\alpha$ & H168  & 210$\pm$37      &  12.7$\pm$1.4 & 15.4$\pm$3.0  & 13             & 0.2\\
G24.79+0.08 Core 1 & H57$\alpha$ & H75   & 1235$\pm$80     &  113.7$\pm$0.80 & 26.2$\pm$2.1  & 44           & 0.4\\
                   & H57$\alpha$ & H168  & 735$\pm$71      &  112.8$\pm$1.3 & 27.2$\pm$3.2  & 25            & 0.4\\
G24.79+0.08 Core 3 & H57$\alpha$ & H75   & 800$\pm$62      &  109.9$\pm$0.58 & 16.4$\pm$1.6  & 46           & 0.6\\
                   & H70$\alpha$ & H168  & 569$\pm$30      &  112.4$\pm$0.45 & 17.3$\pm$1.1  & 31           & 0.2\\
G24.85+0.09 Core 1 & H57$\alpha$ & H75   & 560$\pm$60      &  111.9$\pm$1.2 & 24.1$\pm$3.2  & 22            & 0.4\\ \hline

    \end{tabular}
    \label{tab:rrl_properties}
  \end{center}

\end{table*}

\section{Results}
\label{sec:results}
$\S$~\ref{sub:g316_res} to \ref{sub:g24.85_res} report the 19 to
93\,GHz continuum and RRL detections towards each of the individual
regions. These are compared to both the L07 24\,GHz observations and
original 8\,GHz WBHR98 data to try and uncover the nature of the
continuum sources detected by L07 but not WBHR98. We focus on analysis
of the continuum rather than the RRL data, which will appear in a
subsequent paper. In this work we simply refer to the RRL detections
as further evidence that the continuum arises from free-free emission.

\subsection{G316.81-0.06}
\label{sub:g316_res}
\subsubsection{Previous observations}
As shown in Figure~\ref{fig:g316_cont_maps} (top-left), L07 detected
three sources in the region: two 24\,GHz continuum sources (core 2 at
the methanol maser position, and core 3 offset $\sim$30$\arcsec$
south-west from the maser) and an ammonia filament extending east from
core 2 and the methanol maser position. Despite the strong
($>$0.5\,Jy) continuum towards core 2 at 24\,GHz, WBHR98 only reported
a single 8\,GHz detection towards this region at the position of core
3. To investigate this apparent discrepancy between the L07 and WBHR98
observations, we first re-examined the WBHR98 8\,GHz data.  With poor
uv-coverage, phase stability and strong sidelobe emission, the image
fidelity was not sufficient to rule out an 8\,GHz detection at the
position of core 2. However, the simultaneously observed WBHR98
6.7\,GHz continuum images have much better image fidelity and show a
non-detection to 5\,mJy at the position of core 2 and a 40\,mJy
detection at core 3 with a similar morphology to the 24\,GHz
detection. The apparent discrepancy in the measured flux density of
core 2 between the lower-frequency, very extended array observations
of WBHR98 and the higher-frequency, compact observations of L07
therefore appears to be related to the nature of the source
emission. Based on the new observations, we investigate the nature of
core 2 below.

\subsubsection{G316.81-0.06: Core 2}
\label{subsub:g316_core2}

As shown in Figure~\ref{fig:g316_cont_maps}, we confirm the strong
24\,GHz continuum L07 detection of G316.81-0.06 core 2. At higher
resolution the cm-continuum emission is resolved into two components
separated by $\sim$10$\arcsec$ north-south. At 93\,GHz we find
unresolved continuum emission at the centre of the two cm-continuum
components.

It is clear from the 19, 24 and 35\,GHz continuum images
(Figure~\ref{fig:g316_cont_maps}) and flux densities
(Table~\ref{tab:cont_properties}), that the bulk of the emission in
core 2 is spatially extended. Although the more extended H168 array
configuration 35\,GHz observations resolve the north-south elongation
into at least two components, these only make up a small fraction of
the flux compared to that measured at the same frequency in the more
compact H75 array configuration. 

\begin{figure*}
\begin{center}
\begin{tabular}{cc}
 \includegraphics[width=5.5cm, angle=-90, trim=0 0 -5 0]{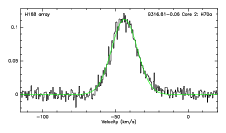} &
 \includegraphics[width=5.5cm, angle=-90, trim=0 0 -5 0]{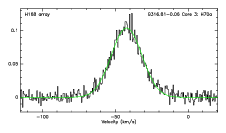} \\
 \includegraphics[width=5.5cm, angle=-90, trim=0 0 -5 0]{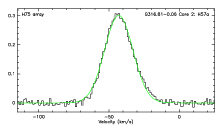} &
 \includegraphics[width=5.5cm, angle=-90, trim=0 0 -5 0]{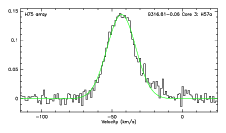} \\
 \includegraphics[width=5.5cm, angle=-90, trim=0 0 -5 0]{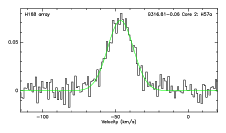} &
 \includegraphics[width=5.5cm, angle=-90, trim=0 0 -5 0]{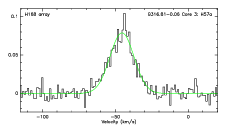} \\
\end{tabular}
\end{center}
\caption{Radio recombination line spectra towards G316.81-0.06 in
         units of Jy/beam. The spectra are extracted from the peak of
         the emission for each of the cores and are labelled by core
         number \& transition in the top right, and array
         configuration in the top left of each spectrum. Overlayed are
         the Gaussian fits to each spectrum outlined in
         $\S$~\ref{sec:obs_red} and used to obtain the values in
         Table~\ref{tab:rrl_properties}.}
\label{fig:g316_spectra}
\end{figure*}

Strong emission from both H70$\alpha$ and H57$\alpha$ are detected
towards core 2 in all the array configurations
(Figure~\ref{fig:g316_spectra}). Similar to the cm-continuum emission,
the peak flux density of the H57$\alpha$ RRL in the more extended
array configuration only recovers a small fraction of the flux density
in the more compact configuration.

To correct for this spatial filtering when creating an SED it is
therefore necessary to sample the source emission at the same spatial
frequencies. Figure~\ref{fig:seds} shows the SED for core 2 sampled
between spatial frequencies of 5$-$10\,k$\lambda$, the mutually
overlapping range between all the 18.9 to 35\,GHz observations. This
shows that on spatial scales of 25-50$\arcsec$, the spectral index is
almost flat -- consistent with optically-thin free-free emission. It
is not possible to unambiguously compare this spectral index with the
flux from the 92 \& 93\,GHz observations due to the smaller primary
beam, much higher spatial frequencies and the potential contribution
from thermal dust emission. However, the similarity between the
92\,GHz H75 array configuration observations and the lower frequency
emission suggests at least some of the 92\,GHz emission may be due to
free-free emission.

The 93\,GHz observations using the more extended H214 array
configuration show a very different structure, with only a single,
weak, unresolved detection directly in the middle of the two compact
lower frequency knots. This source is also coincident with an
N$_2$H$^+$ core (Longmore et al. in prep) tracing dense molecular gas,
so we postulate this 93GHz emission is likely to be from thermal dust.
Although the emission from this source is probably distinct from that
of core 2 at lower frequencies, they are spatially associated within
the beam of the lower resolution data, and as such are both labelled
as `core 2'.

Finally, there is also a strong 93\,GHz continuum source at the
position of the ammonia peak, core 1. The spectral index inferred from
the upper limits of the lower frequency data at this position means
this is also likely to be thermal dust emission. As this source is
substantially outside of the primary beam, the flux measurement
reported in Table~\ref{tab:cont_properties} is highly uncertain.

\subsection{G331.28-0.19}
\label{sub:g331_res}
\begin{figure*}
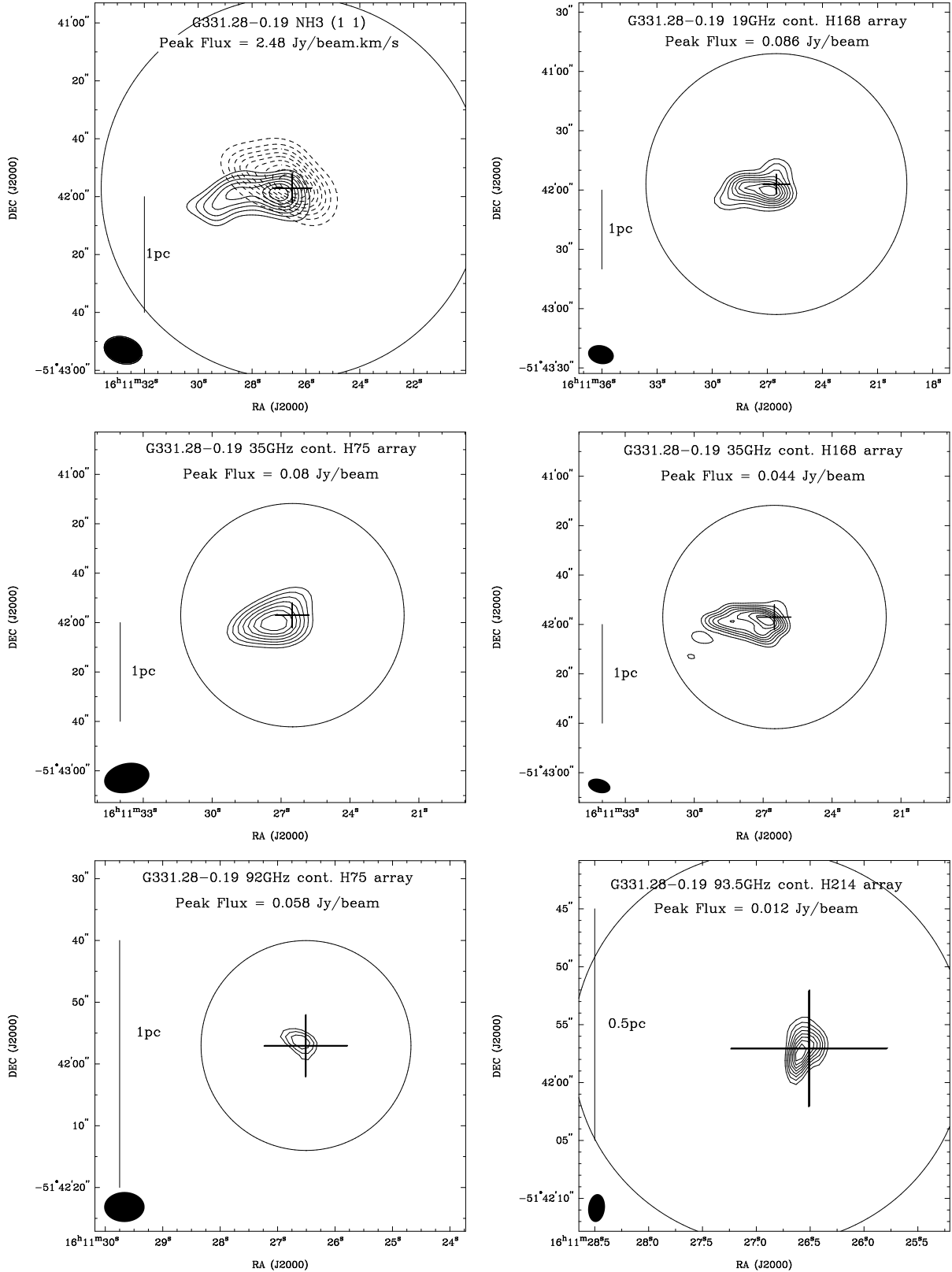

\begin{center}
\begin{tabular}{cc}
 \includegraphics[width=7.0cm, angle=-90, trim=0 0 -5 0]{figs/postscripts/g331/g331.28.1_1.mom0_2.ps} &
 \includegraphics[width=7.0cm, angle=-90, trim=0 0 -5 0]{figs/postscripts/g331/g331.18775.cont_h168.ps} \\
 \includegraphics[width=7.0cm, angle=-90, trim=0 0 -5 0]{figs/postscripts/g331/g331.34586.cont_h75.ps} &
 \includegraphics[width=7.0cm, angle=-90, trim=0 0 -5 0]{figs/postscripts/g331/g331.34604.cont_h168.ps} \\
 \includegraphics[width=7.0cm, angle=-90, trim=0 0 -5 0]{figs/postscripts/g331/g331.92056.cont_h75.ps} &
 \includegraphics[width=7.0cm, angle=-90, trim=0 0 -5 0]{figs/postscripts/g331/g331.93500.cont_h214.ps} \\
\end{tabular}
\end{center}
\caption{G331.28-0.19 maps. The top-left image shows the integrated
  $\nhone$ emission as dashed contours and the 24GHz continuum
  emission as solid contours (taken from L07). The remaining images
  show the continuum emission at each frequency and array
  configuration. The peak flux shown at the top of each image
  corresponds to the maximum value in that image rather than the fits
  to the emission reported in
  Table~\ref{tab:cont_properties}. Contours are in 10\% steps of the
  peak value with the maximum contour at 90\%. The primary and
  synthesised beam sizes are shown as a circle and filled ellipse,
  respectively. The linear scale is illustrated by the line in the
  lower left of each image. The position of the methanol maser
  emission is shown as crosses, with the same angular size in each
  image.}
\label{fig:g331_cont_maps}
\end{figure*}

Figure~\ref{fig:g331_cont_maps} shows images of the continuum emission
detected towards the region. The morphology of the 19 and 35\,GHz
continuum emission is very similar to that observed at 24\,GHz by L07:
a single source is seen peaking close to the methanol maser position
with an extended tail towards the east. This continuum emission is
offset to the south east from an L07 ammonia core which is extended
north-east to south-west.

WBHR98 reported no 8\,GHz continuum source in the
region. Re-examination of the WBHR98 8\,GHz data shows a potential
source of 4mJy at this position with a similar morphology. However,
due to poor weather, only 2$\times$2 minute cuts were observed in the
WBHR98 observations so the image fidelity is poor and the emission
brightness lay below their detection criteria. However,
\citet{phillips1998} report weak ($\sim$3.5\,mJy) continuum emission
at 8\,GHz, extended east-west, with a similar morphology to that in
Figure~\ref{fig:g331_cont_maps}.

\begin{figure*}
\begin{center}
\begin{tabular}{cc}
 \includegraphics[width=5.5cm, angle=-90, trim=0 0 -5 0]{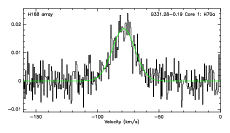} &
 \includegraphics[width=5.5cm, angle=-90, trim=0 0 -5 0]{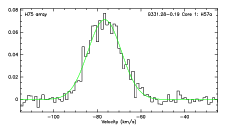} \\
\end{tabular}
\end{center}
\caption{Radio recombination line spectra towards G331.28-0.19 in
         units of Jy/beam. The spectra are extracted from the peak of
         the emission and are labelled by core number \& transition in
         the top right, and array configuration in the top left of
         each spectrum. Overlayed are the Gaussian fits to each
         spectrum outlined in $\S$~\ref{sec:obs_red} and used to obtain
         the values in Table~\ref{tab:rrl_properties}.}
\label{fig:g331_spectra}
\end{figure*}

The H70$\alpha$ and H57$\alpha$ RRL emission detected towards the
continuum (Figure~\ref{fig:g331_spectra}) and spectral index of
$-$0.07 derived from the SED (Figure~\ref{fig:seds}) confirms the
continuum between 19 and 35\,GHz as free-free emission.

A compact continuum source is also detected at 92 and 93\,GHz. This is
offset from the peak of the lower frequency emission and is instead
associated with the methanol maser emission, the peak of the L07
$\nhone$ ammonia emission and the unresolved $\nhfour$ and (5,5)
emission. Due to a combination of the lower sensitivity to extended
emission of these observations and the lower optical depth of
free-free emission at higher frequencies, the 92 \& 93\,GHz emission
could be from the densest and brightest free-free component inside the
larger region traced by the lower frequency emission. However, given
the similar synthesised beam sizes of the 35\,GHz H168 and 92\,GHz H75
observations, it is difficult to understand the offset in the peak of
the emission. We favour an alternative explanation in which the 92 \&
93 GHz emission is dominated by thermal dust emission rather than
free-free emission.

In summary, G331.28-0.19 core 1 appears to be an optically thin HII
region.  The 92 GHz emission and the ammonia emission are likely
tracing a hot molecular core located at the edge of the HII region,
similar to the case of G29.96-0.02 \citep{cesaroni1998}.  The 8 GHz
non-detection by WBHR98 can be attributed to their poorer sensitivity
for this source compared to the rest of their sample.

\subsection{G12.68-0.18}
\label{sub:g12.68_res}

\begin{figure*}
\begin{center}
\begin{tabular}{cc}
 \includegraphics[width=5.5cm, angle=-90, trim=0 0 -5 0]{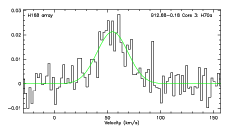} &
 \includegraphics[width=5.5cm, angle=-90, trim=0 0 -5 0]{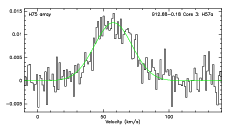} \\
 \includegraphics[width=5.5cm, angle=-90, trim=0 0 -5 0]{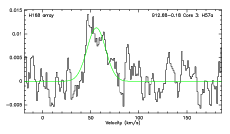} &
\end{tabular}
\end{center}
\caption{Radio recombination line spectra towards G12.68-0.18 in units of 
         Jy/beam. The spectra are extracted from the peak of the
         emission and are labelled by core number \& transition in the
         top right, and array configuration in the top left of each
         spectrum. Overlayed are the Gaussian fits to each spectrum
         outlined in $\S$~\ref{sec:obs_red} and used to obtain the
         values in Table~\ref{tab:rrl_properties}.}
\label{fig:g12.68_spectra}
\end{figure*}

Figure~\ref{fig:g12.68_cont_maps} shows images of the continuum
emission detected towards the region. The morphology of the 19 and
35\,GHz continuum emission is similar to that observed at 24\,GHz by
L07: a single extended source is seen 10$\arcsec$ north-east of the
methanol maser emission and offset from the ammonia cores in the
region. Figure~\ref{fig:g12.68_spectra} shows the H57$\alpha$ and
H70$\alpha$ RRL spectra detected towards the peak of the continuum
emission, confirming this arises from ionised
gas. Figure~\ref{fig:seds} shows the SED for this source. Although the
quality of the fit is not particularly good, the spectral index of
$-0.4\pm0.11$ is clearly inconsistent with optically thick free-free
or dust emission. The extremely negative spectral index is biased by
the low flux density of the 35\,GHz H168 data point, which is probably
due to missing flux. Removing this point before fitting results in a
spectral index of $-0.18\pm0.09$.

Re-inspection of the WBHR98 data shows this source is
$\sim$10$\arcmin$ from the WBHR98 pointing centre, significantly
outside their half-power beamwidth of 5.7$\arcmin$. Any emission would
therefore have suffered from primary beam attenuation in their
observations. 

Figure~\ref{fig:g12.68_cont_maps} also shows a second, unresolved
source detected at 93\,GHz, 5$\arcsec$ south of the maser position,
coincident with the unresolved $\nhone \rightarrow$ (5,5) core
reported in L07 (core 4). The non-detection at frequencies $<$93\,GHz
makes it likely that the emission mechanism for this source is from
dust rather than thermal free-free emission. 

In summary, G12.68-0.18 core 3 is an extended region of optically-thin
ionised gas that was not detected by WBHR98 as it lies far outside
their primary beam. 

\subsection{G19.47+0.17}
\label{sub:g19.47_res}

\begin{figure*}
\begin{center}
\begin{tabular}{cc}
 \includegraphics[width=5.5cm, angle=-90, trim=0 0 -5 0]{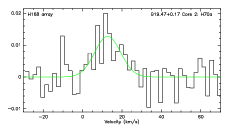} &
 \includegraphics[width=5.5cm, angle=-90, trim=0 0 -5 0]{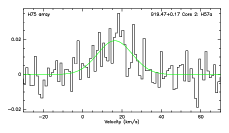} \\
\end{tabular}
\end{center}
\caption{Radio recombination line spectra towards G19.47+0.17 in units
         of Jy/beam. The spectra are extracted from the peak of the
         emission and are labelled by core number \& transition in the
         top right, and array configuration in the top left of each
         spectrum. Overlayed are the Gaussian fits to each spectrum
         outlined in $\S$~\ref{sec:obs_red} and used to obtain the
         values in Table~\ref{tab:rrl_properties}.}
\label{fig:g19.47_spectra}
\end{figure*}

Figure~\ref{fig:g19.47_cont_maps} shows images of the continuum
emission detected towards the region. The morphology of the 19 and
35\,GHz continuum emission is similar to that observed at 24\,GHz by
L07: an unresolved continuum source, labelled core 2, lies
$\sim$15$\arcsec$ north-west of the methanol maser
position. H57$\alpha$ and H70$\alpha$ RRL emission is detected towards
the peak of the continuum emission (Figure~\ref{fig:g19.47_spectra}),
confirming the emission arises from ionised gas. The spectral index
(Figure~\ref{fig:seds}) is consistent with optically-thin free-free
emission.

No 8\,GHz continuum emission was reported at this location by
WBHR98. Re-examination of the WBHR98 data shows this region lies
outside the primary beam so the continuum sensitivity would have
suffered significantly from primary beam attenuation.

A second, weaker source is seen only at 92 and 93\,GHz within a few
arc-seconds of the methanol maser position and unresolved $\nhone
\rightarrow (5,5)$ emission, designated core 1 in L07. At 35\,GHz the
3$\sigma$ upper limit is 7\,mJy. This implies a spectral index between
35 and 92\,GHz of $\alpha >2$, which rules out either optically-thick
or thin bremsstrahlung emission. We conclude that the core 1 emission
is most likely from warm dust.

\subsection{G24.79+0.08}
\label{sub:g24.79_res}

\begin{figure*}
\begin{center}
\begin{tabular}{cc}
 \includegraphics[width=5.5cm, angle=-90, trim=0 0 -5 0]{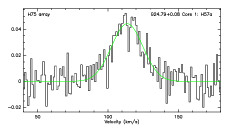} &
 \includegraphics[width=5.5cm, angle=-90, trim=0 0 -5 0]{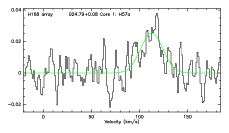} \\
 \includegraphics[width=5.5cm, angle=-90, trim=0 0 -5 0]{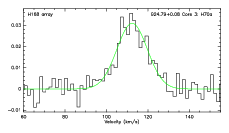} &
 \includegraphics[width=5.5cm, angle=-90, trim=0 0 -5 0]{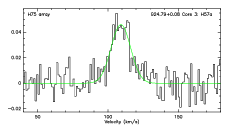} \\
\end{tabular}
\end{center}
\caption{Radio recombination line spectra towards G24.79+0.08 in units of 
         Jy/beam. The spectra are extracted from the peak of the
         emission for each of the cores and are labelled by core
         number \& transition in the top right, and array
         configuration in the top left of each spectrum. Overlayed are
         the Gaussian fits to each spectrum outlined in
         $\S$~\ref{sec:obs_red} and used to obtain the values in
         Table~\ref{tab:rrl_properties}.}
\label{fig:g24.79_spectra}
\end{figure*}

Figure~\ref{fig:g24.79_cont_maps} shows images of the continuum
emission detected towards the region. The morphology at 19 and 35\,GHz
is very similar to that observed at 24\,GHz by L07: an unresolved
source at the methanol maser position (core 1) and a slightly extended
source at the edge of the primary beam (core 3). Although, core 3 was
not detected at 92 \& 93\,GHz, it lies outside the primary beam at
$>$24\,GHz making both detection more difficult and the flux
determination more uncertain. Figure~\ref{fig:g24.79_spectra}, shows
H57$\alpha$ and H70$\alpha$ RRL emission is detected at the peak of
the continuum emission towards core 3 and H57$\alpha$ is detected
towards core 1, confirming both these sources contain ionised gas.

WBHR98 reported a 1$\sigma$ non-detection to 3mJy at both 24\,GHz
continuum positions. However, \citet{forstercaswell2000} reported two
8.6\,GHz continuum sources in the proximity of the methanol maser
emission, with flux densities of 36 and 51\,mJy. Re-examining the
calibrated WBHR98 $uv$ data shows residual phase errors in the data
for this region. The image shows 8\,GHz continuum emission in the
region but its location is unclear, resulting in the reported
non-detection.

Core 1 is well known in the literature as a hyper-compact \hii region
\citep[e.g.][]{beltran2007}. Figure~\ref{fig:seds} shows the SED at
the frequencies observed in this work which are consistent with the
source transitioning between optically-thick and thin free-free
emission. The single-component spectral index attributed to this
source in Figure~\ref{fig:seds} is therefore not a good representative
of the true SED.

In summary, these observations confirm both the L07 24\,GHz continuum
detections and show core 3 is an extended region of optically-thin
ionised gas which was missed by WBHR98. Core 1 is a well known
optically-thick, hyper-compact \hii~region which was missed by WBHR98
due a combination of the steeply rising flux density between 8 and
24\,GHz and calibration problems.

\subsection{G24.85+0.09}
\label{sub:g24.85_res}
\begin{figure*}
\begin{center}
\begin{tabular}{cc}
 \includegraphics[width=5.5cm, angle=-90, trim=0 0 -5 0]{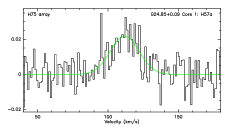} &
\end{tabular}
\end{center}
\caption{Radio recombination line spectrum towards G24.85+0.09 in units of 
         Jy/beam. The spectrum is extracted from the peak of the
         emission and labelled by core number \& transition in the top
         right, and array configuration in the top left of the
         spectrum. Overlayed are the Gaussian fits to each spectrum
         outlined in $\S$~\ref{sec:obs_red} and used to obtain the
         values in Table~\ref{tab:rrl_properties}.}
\label{fig:g24.85_spectra}
\end{figure*}

Figure~\ref{fig:g24.85_cont_maps} shows images of the continuum
emission detected towards the region. The morphology at 19 and 35\,GHz
is similar to that observed at 24\,GHz by L07: a compact source is
seen peaking $\sim$5$\arcsec$ west of the methanol maser position,
with a fainter tail extending towards the south-east. No emission was
detected at 92 or 93\,GHz. The detection of H57$\alpha$
(Figure~\ref{fig:g24.85_spectra}) confirms the presence of free-free
emission towards the source. Figure~\ref{fig:seds} shows the SED. The
spectral index of 0.4 determined from the fit is not very robust due
to the flux uncertainties and the narrow range of detected
frequencies. Nevertheless, the spectral index is sufficient to rule
out dust as the primary contributor to the emission. The spectral
index of 0.4 is intermediate between the canonical values of $+$2 and
$-$0.1, suggesting that the sampled frequencies are near the turn over
region. Re-examination of the WBHR98 data shows a 3$\sigma$
non-detection to 6mJy at 8\,GHz. Extrapolating the WBHR98 upper limit
assuming a spectral index of $+$2 results in a 19\,GHz flux density
well below the measured value (see Figure~\ref{fig:seds}). From this,
and the extended source size we assume the emission is optically thin.

\begin{table}
  \caption{Nature of the L07A and newly detected continuum sources
    towards each of the regions derived from the spectral energy
    distributions (see $\S$\ref{sub:seds}). }
  \label{tab:cont_nature}
  \begin{tabular}{|c|c|l|c|} \hline \hline
    
    Region        & Core & Nature of emission        & Freq.\\\hline
    G316.81-0.06  & 2    & optically thin free-free  & 24\,GHz    \\
    G331.28-0.19  & 1    & optically thin free-free  & 24\,GHz     \\
    G12.68-0.18   & 3    & optically thin free-free  &  24\,GHz    \\
    G12.68-0.18   & 4    & dust                      & 92\,GHz\\
    G19.47+0.17   & 1    & dust                      & 92\,GHz\\
    G19.47+0.17   & 2    & optically thin free-free  &  24\,GHz     \\
    G24.79+0.08   & 1    & optically thick free-free  &  24\,GHz     \\
    G24.85+0.09   & 1    & optically thin free-free  &  24\,GHz     \\ \hline
        
  \end{tabular}
\end{table}

\subsection{Spectral Energy Distributions}
\label{sub:seds}

\begin{figure*}
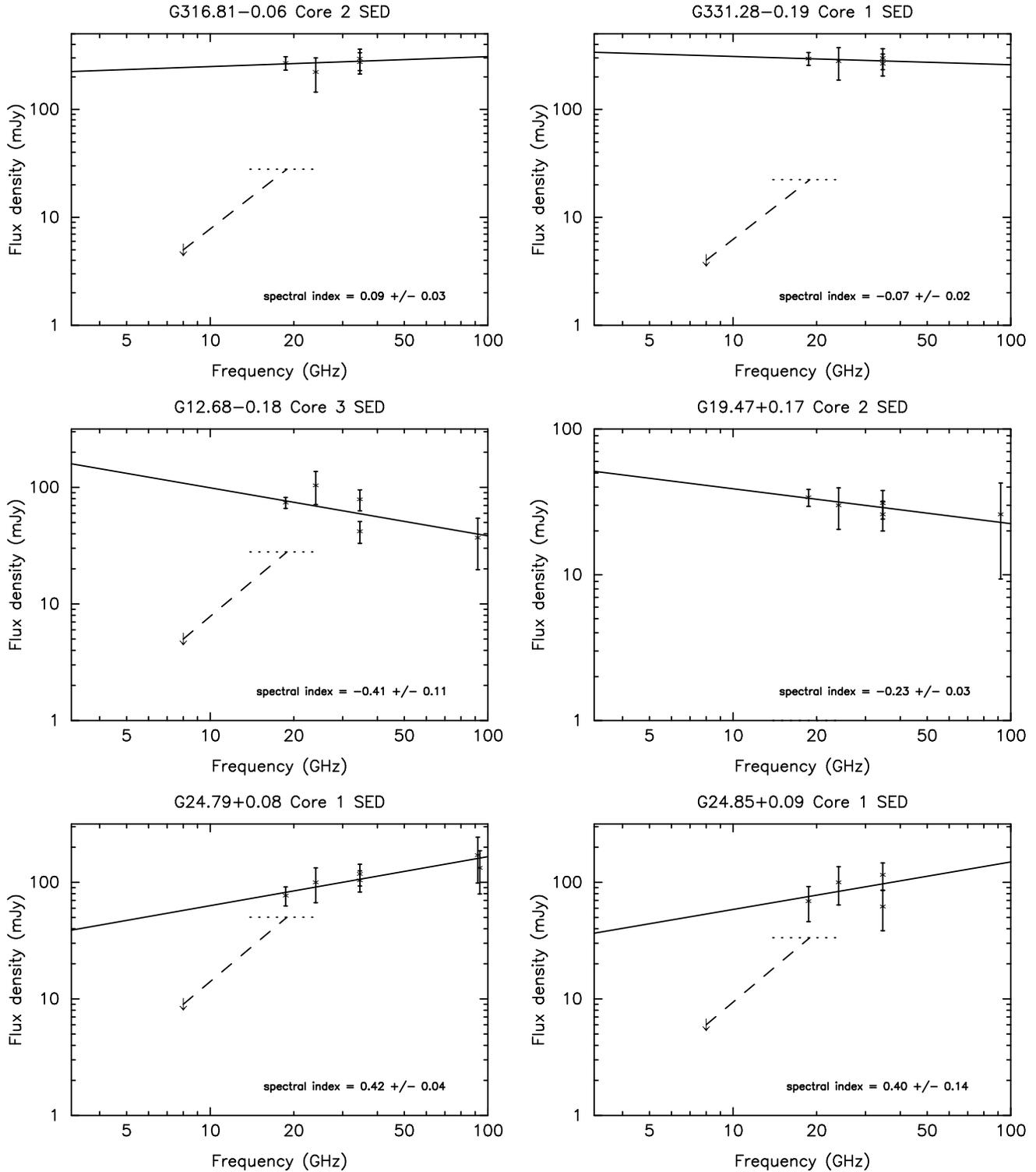

\begin{center}
\begin{tabular}{cc}
 \includegraphics[width=6.5cm, angle=-90, trim=0 0 -5 0]{figs/seds/g316_core2_sed_uvfit_uvrange5-10.ps} &
 \includegraphics[width=6.5cm, angle=-90, trim=0 0 -5 0]{figs/seds/g331_core1_sed_uvfit_err.ps} \\
 \includegraphics[width=6.5cm, angle=-90, trim=0 0 -5 0]{figs/seds/g12.68_core3_sed_uvfit_err.ps} &
 \includegraphics[width=6.5cm, angle=-90, trim=0 0 -5 0]{figs/seds/g19.47_core2_sed_uvfit_err.ps} \\
 \includegraphics[width=6.5cm, angle=-90, trim=0 0 -5 0]{figs/seds/g24.79_core1_sed_uvfit_err.ps} &
 \includegraphics[width=6.5cm, angle=-90, trim=0 0 -5 0]{figs/seds/g24.85_core1_sed_uvfit_err.ps} \\

\end{tabular}
\end{center}
\caption{Spectral energy distributions for each of the sources
  detected at 24\,GHz and not 8\,GHz in WBHR98. Crosses and error bars
  show the flux densities and their corresponding uncertainties
  measured in this work and L07A. To mitigate problems comparing
  fluxes of extended sources sampled at different spatial frequencies,
  extended sources were re-imaged only including visibilities in the
  mutually overlapping spatial frequency range. Fluxes extracted from
  these images therefore sample the sources over similar spatial
  scales. The solid line shows the weighted best-fit to the detections
  and the corresponding spectral index is given as text. The arrows
  show the detection limit at 8\,GHz reported in WBHR98.  The dashed
  line shows the SED between 8 to 19\,GHz of a source with a flux
  density at 8\,GHz equal to the lower limit reported by WBHR98 and
  spectral index of 2. The horizontal dotted line shows the expected
  19GHz flux that would be expected for a source of optically thick
  free-free emission, which lies just below the detection threshold of
  WBHR98. The 8\,GHz detection limit and dashed line are omitted for
  G19.47+0.17 as this was too far outside the WBHR98 primary beam to
  determine a reliable detection limit. The dashed lines in the
  remaining sources show that, with the exception of G24.79+0.08, the
  measured flux densities in this work can not have arisen solely from
  an optically thick source of free-free emission lying just below the
  WBHR98 detection limit. }
\label{fig:seds}
\end{figure*}

Figure~\ref{fig:seds} shows the SEDs towards each of the cores
detected at 24\,GHz by L07 and not at 8\,GHz by WBHR98. Our aim was
to distinguish between the different expected emission mechanisms:
optically-thin/thick thermal free-free emission ($\alpha \sim -$0.1 or
2, respectively. Given the different range of spatial frequencies
probed between the different observations, the observations at
$<$90\,GHz were re-imaged selecting only visibilities in the mutually
overlapping spatial frequency range. In practice this meant dropping
the longest baselines at higher frequency, thus biasing the SED
towards the large scale structure. As a result it was not possible to
resolve G316.81-0.06 Core 2 into the two separate components, and only
a composite SED is given. After re-imaging and re-fitting the emission
to determine the flux density (as outlined in $\S$~\ref{sec:obs_red}),
the spectral index was calculated using a weighted, linear
least-squares fit. Uncertainties in the individual flux density
measurements were calculated by adding in quadrature the formal error
of the fit to the emission and the estimated uncertainty in absolute
flux (see $\S$~\ref{sec:obs_red}). The absolute flux uncertainty was
dominant in nearly all cases. As there were no overlapping spatial
frequencies between the higher resolution observations at 92/93.5\,GHz
and the remaining data, it was not possible to re-image these data in
the same way. As a result, the 92/93.5\,GHz observations miss much of
the flux seen towards extended sources at lower frequencies,
potentially biasing the spectral index. The uncertainty to the
spectral index is provided in each of the figures.

Table~\ref{tab:cont_nature} shows the conclusions as to the nature of
the 24\,GHz continuum emission in L07 based on the SED
fitting. Despite the sometimes large uncertainty in spectral index,
the SEDs are sufficient to rule out an optically-thick component as
the dominant emission mechanism at this spatial scale in
G316.81$-$0.02 core 2, G331.28$-$0.19 core 1, G12.68$-$0.18 core 3 and
G19.47+0.17 core 2. This does not preclude the existence of compact
ionised gas components at smaller spatial scales (indeed, density
gradients may well exist), only that at this position and spatial
scale diffuse, optically thin ionised gas is dominant.

So how do these SEDs compare to the WBHR98 data? Where applicable, the
WBHR98 8\,GHz upper limits are shown as arrows in
Figure~\ref{fig:seds}. Given the sometimes several orders of magnitude
discrepancy between the expected 8\,GHz flux extrapolated from the
SEDs, the non-detections seem initially inconsistent. One potential
explanation may be that the emission is optically-thick between 8 and
19\,GHz, and as a result drops below the WBHR98 detection limit. To
test this we have extrapolated the expected SED between 8 and 19\,GHz
assuming the emission is optically thick and just falls below the
WBHR98 detection limit (dashed line in Figure~\ref{fig:seds}). While
this may be a viable explanation for G24.79+0.08 core 1 and possibly
G24.85+0.09, it clearly fails for the remaining sources. We conclude
the only other viable explanation is that WBHR98 observations must
have missed the spatially extended optically-thin emission through
either spatial filtering, reduced surface brightness sensitivity or a
combination of the two.

\section{Discussion: the ``missing'' cm-continuum sources}
\label{sec:discussion_missing_sources}

Observations of cm-continuum emission towards regions of massive star
formation provide a powerful indicator of the region's
age. Simplistically, detection of relatively weak (1-5\,mJy) emission
from shock-ionised gas signals the existence of a powering source
(that may be driving a thermal jet for example), while stronger
emission from photo-ionised gas signals a young massive stellar object
has evolved sufficiently to have begun ionising its surrounding
environment. Conversely, lack of bright cm-continuum emission is often
used to infer the region is at an evolutionary stage prior to the
development of an UC\hii region. As such, previous large,
high-resolution, cm-continuum surveys have proven invaluable in
helping to assess the relative ages of many massive star formation
regions. However, from our original sample of 21 massive star
formation regions, selected from such a survey to contain no
cm-continuum emission, we have confirmed that 6 of these are in fact
associated with cm-continuum emission. In
$\S$~\ref{sub:reason_missing_sources} we summarise the reasons for
these discrepancies for the sources we have studied in this work and
in $\S$~\ref{sub:implications} discuss the implications this has on
the implied evolutionary stage of the objects. We then assess how
these results may affect the interpretation of the whole WBHR98
sample.

\subsection{Reasons for the ``missing' cm-continuum sources}
\label{sub:reason_missing_sources}
We find three reasons why the cm-continuum sources in this work were
not detected at 8\,GHz by WBHR98. These can be separated by the nature
of the source emission and the sensitivity limit of the survey as
follows:

\begin{enumerate}

\item the free-free emission is dominated by compact components, is
      optically-thick at 8\,GHz and lies below the WBHR98 detection
      limit (i.e. G24.79+0.08 core 1)

\item the emission is significantly above the WBHR98 detection limit
  but much more spatially extended than the WBHR98 beam
  (i.e. G316.81-0.06, G331.28-0.19 and G24.85+0.09)

\item the emission lies well outside the primary beam of WBHR98
      (i.e. G12.68-0.18 and G19.47+0.17)

\end{enumerate}

In each of the three cases, the reason for the non-detection at 8\,GHz
is straightforward: (i) is due to the sensitivity limits of WBHR98;
(ii) is due to the WBHR98 observations resolving-out extended
emission, and, (iii) is due to the significantly reduced sensitivity
of the observations outside of the primary beam.

\subsection{Implications}
\label{sub:implications}

\subsubsection{Methanol masers and an evolutionary sequence}
\label{subsub:imp_meth_evol}

Based on the WBHR98 data, 15 of the 21 regions in the initial L07
survey were selected as being methanol maser sites devoid of 8\,GHz
continuum emission and were thus assumed to be in an evolutionary
stage prior to UC\hii region formation. The results from this work
show that in fact only 9 of the original 21 regions contain no
free-free emission and the remaining 12 are either at the HC\hii
region stage [case (i)] or in fact substantially older [cases (ii or
  iii)].

WBHR98 found $\sim$80\% of their methanol maser sites were not
associated with radio continuum emission and suggested methanol masers
may form before an UCH{\scriptsize II}~region develops. While this
work does not contradict the conclusions of WBHR98 - methanol masers
do trace very early stages of massive star formation - our results
suggest a substantially larger fraction of methanol masers than
determined by WBHR98 are associated with more evolved stages,
UCH{\scriptsize II}~regions and later. However, it is not possible to
rule out methanol masers as signposts of the pre-HII phase. Given the
clustered mode of massive star formation, these youngest sources may
simply be coincident with more evolved massive young stellar objects.

Our findings support the idea that methanol masers commonly occur in
both a very early phase before a detectable UCH{\scriptsize II}~region
develops (this may include the hypercompact H{\scriptsize II}~region
phase), and after the formation of a UCH{\scriptsize II}~region where
the ionised gas is expanding and becomes extended enough that it may
be missed by extended interferometer configurations, such as that used
by WBHR98, due to spatial filtering/poor surface brightness
sensitivity.

In order to fully understand the relative occurrence of methanol maser
sites in these three stages, we suggest a radio continuum survey that
is sensitive to more extended radio continuum emission be
undertaken. Such a survey should observe a large number of methanol
maser sites without associated radio continuum emission, as identified
by WBHR98.

\subsubsection{Dust masses derived from sub-mm flux densities}
\label{subsub:imp_seds}

Inadvertently ignoring a potentially substantial flux contribution (up
to $\sim$0.5\,Jy) from free-free emission has implications for
deriving dust masses from sub-mm flux densities. Given the sharply
rising flux density of dust with frequency this overestimate will be
most pronounced at lower frequencies. Comparison with sub-mm flux
densities towards massive star formation regions
\citep[e.g.][]{hill2005}, shows that in the worse case scenario,
missing a 0.5\,Jy contribution from free-free emission will result in
an over-estimate of dust mass by a factor of $\sim$2.

\section{Conclusions}
\label{sec:conclusion}

We have presented 19 to 93\,GHz continuum and H70$\alpha$ $+$
H57$\alpha$ radio recombination line observations with the Australia
Telescope Compact Array towards 6 hot molecular cores (HMCs)
associated with methanol maser emission. We confirm the results of
\citet{L07A} that previous continuum surveys have missed several,
sometimes bright ($\sim$0.5Jy) cm-continuum sources. If the original
\citet{L07A} sample of 21 massive star formation regions are
representative of the population as a whole, the evolutionary stage of
a large number of regions have been mis-classified. Rather than being
very young objects prior to UC\hii region formation, they are, in
fact, associated with free-free emission and thus significantly older.

In addition, inadvertently ignoring a potentially substantial flux
contribution (up to $\sim$0.5\,Jy) from free-free emission has
implications for dust masses derived from sub-mm flux densities,
particularly the larger spatial scales probed by single-dish
telescopes. Dust masses may be overestimated by a factor of $\sim$2.

We note that cm-continuum data from observations simultaneously
studying masers are particularly susceptible to this missing flux
problem, due to the extended array configurations needed to achieve
sufficient angular resolution. The data in this work adds further
evidence that high resolution interferometric continuum data from such
data may not be telling the whole picture.

\section{Acknowledgements}
We thank Paul Ho and the referee, Stuart Lumsden, for instructive
comments. The Australia Telescope is funded by the Commonwealth of
Australia for operation as a National Facility managed by CSIRO.

\bibliography{snl_atca_rrl_2}

\begin{figure*}
\begin{center}
\begin{tabular}{cc}
 \includegraphics[width=7cm, angle=-90, trim=0 0 -5 0]{figs/postscripts/g316/g316.81.1_1.mom0_2.ps} &
 \includegraphics[width=7cm, angle=-90, trim=0 0 -5 0]{figs/postscripts/g316/g316.18775.cont_h168.ps} \\
 \includegraphics[width=7cm, angle=-90, trim=0 0 -5 0]{figs/postscripts/g316/g316.34586.cont_h75.ps} &
 \includegraphics[width=7cm, angle=-90, trim=0 0 -5 0]{figs/postscripts/g316/g316.34604.cont_h168.ps} \\
 \includegraphics[width=7cm, angle=-90, trim=0 0 -5 0]{figs/postscripts/g316/g316.92041.cont_h75.ps} &
 \includegraphics[width=7cm, angle=-90, trim=0 0 -5 0]{figs/postscripts/g316/g316.93500.cont_h214.ps} \\
\end{tabular}
\end{center}
\caption{G316.81-0.06 maps. The top-left image shows the integrated
  $\nhone$ emission as dashed contours and the 24GHz continuum
  emission as solid contours (taken from L07).  The numbers show the
  labels given to each of the cores. The remaining images show the
  continuum emission at each frequency and array configuration. The
  peak flux shown at the top of each image corresponds to the maximum
  value in that image rather than the fits to the emission reported in
  Table~\ref{tab:cont_properties}. Contours are in 10\% steps of the
  peak value with the maximum contour at 90\%. The primary and
  synthesised beam sizes are shown as a circle and filled ellipse,
  respectively. The linear scale is illustrated by the line in each
  image. The position of the methanol maser emission and 8\,GHz
  continuum from WBHR98 are shown as crosses and a box, respectively
  in all images and have the same angular size in each image.}
\label{fig:g316_cont_maps}
\end{figure*}

\begin{figure*}
\begin{center}
\begin{tabular}{cc}
 \includegraphics[width=6.8cm, angle=-90, trim=0 0 -5 0]{figs/postscripts/g12.68/g12.68.1_1.mom0_2.ps} &
 \includegraphics[width=6.8cm, angle=-90, trim=0 0 -5 0]{figs/postscripts/g12.68/g12.68-0.18.18768.cont_h168.ps} \\
 \includegraphics[width=6.8cm, angle=-90, trim=0 0 -5 0]{figs/postscripts/g12.68/g12.68-0.18.34586.cont_h75.ps} &
 \includegraphics[width=6.8cm, angle=-90, trim=0 0 -5 0]{figs/postscripts/g12.68/g12.68-0.18.34592.cont_h168.ps} \\
 \includegraphics[width=6.8cm, angle=-90, trim=0 0 -5 0]{figs/postscripts/g12.68/g12.68-0.18.92018.cont_h75.ps} &
 \includegraphics[width=6.8cm, angle=-90, trim=0 0 -5 0]{figs/postscripts/g12.68/g12.68-0.18.93500.cont_h214.ps} \\
 \\
\end{tabular}
\end{center}
\caption{G12.68-0.18 maps. The top-left image shows the integrated
  $\nhone$ emission as dashed contours and the 24GHz continuum
  emission as solid contours (taken from L07).  The numbers show the
  labels given to each of the cores. The remaining images show the
  continuum emission at each frequency and array configuration. The
  peak flux shown at the top of each image corresponds to the maximum
  value in that image rather than the fits to the emission reported in
  Table~\ref{tab:cont_properties}. Contours are in 10\% steps of the
  peak value with the maximum contour at 90\%. The primary and
  synthesised beam sizes are shown as a circle and filled ellipse,
  respectively. The linear scale is illustrated by the line in the
  lower left of each image. The position of the methanol maser
  emission is shown as crosses, with the same angular size in each
  image. }
\label{fig:g12.68_cont_maps}
\end{figure*}

\clearpage

\begin{figure*}
\begin{center}
\begin{tabular}{cc}
 \includegraphics[width=7.0cm, angle=-90, trim=0 0 -5 0]{figs/postscripts/g19.47/g19.47.1_1.mom0_2.ps} &
 \includegraphics[width=7.0cm, angle=-90, trim=0 0 -5 0]{figs/postscripts/g19.47/g19.47.18768.cont_h168.ps} \\
 \includegraphics[width=7.0cm, angle=-90, trim=0 0 -5 0]{figs/postscripts/g19.47/g19.47.34586.cont_h75.ps} &
 \includegraphics[width=7.0cm, angle=-90, trim=0 0 -5 0]{figs/postscripts/g19.47/g19.47.34604.cont_h168.ps} \\
 \includegraphics[width=7.0cm, angle=-90, trim=0 0 -5 0]{figs/postscripts/g19.47/g19.47.92018.cont_h75.ps} &
 \includegraphics[width=7.0cm, angle=-90, trim=0 0 -5 0]{figs/postscripts/g19.47/g19.47.93500.cont_h214.ps} \\
\end{tabular}
\end{center}
\caption{G19.47+0.17 maps. The top-left image shows the integrated
  $\nhone$ emission as dashed contours and the 24GHz continuum
  emission as solid contours (taken from L07).  The numbers show the
  labels given to each of the cores. The remaining images show the
  continuum emission at each frequency and array configuration. The
  peak flux shown at the top of each image corresponds to the maximum
  value in that image rather than the fits to the emission reported in
  Table~\ref{tab:cont_properties}. Contours are in 10\% steps of the
  peak value with the maximum contour at 90\%. The primary and
  synthesised beam sizes are shown as a circle and filled ellipse,
  respectively. The linear scale is illustrated by the line in the
  lower left of each image. The position of the methanol maser
  emission is shown as a cross, with the same angular size in each
  image.}
\label{fig:g19.47_cont_maps}
\end{figure*}

\begin{figure*}
\begin{center}
\begin{tabular}{cc}
 \includegraphics[width=7.0cm, angle=-90, trim=0 0 -5 0]{figs/postscripts/g24.79/g24.79.1_1.mom0_2.ps} &
\includegraphics[width=7.0cm, angle=-90, trim=0 0 -5 0]{figs/postscripts/g24.79/g24.79+0.08.18768.cont_h168.ps} \\
 \includegraphics[width=7.0cm, angle=-90, trim=0 0 -5 0]{figs/postscripts/g24.79/g24.79+0.08.34586.cont_h75.ps} &
 \includegraphics[width=7.0cm, angle=-90, trim=0 0 -5 0]{figs/postscripts/g24.79/g24.79+0.08.34592_h168_cont.ps} \\
 \includegraphics[width=7.0cm, angle=-90, trim=0 0 -5 0]{figs/postscripts/g24.79/g24.79+0.08.91997.cont_h75.ps} &
 \includegraphics[width=7.0cm, angle=-90, trim=0 0 -5 0]{figs/postscripts/g24.79/g24.79+0.08.93500.cont_h214.ps} \\
\end{tabular}
\end{center}
\caption{G24.79+0.08 maps. The top-left image shows the integrated
  $\nhone$ emission as dashed contours and the 24GHz continuum
  emission as solid contours (taken from L07).  The numbers show the
  labels given to each of the cores. The remaining images show the
  continuum emission at each frequency and array configuration. The
  peak flux shown at the top of each image corresponds to the maximum
  value in that image rather than the fits to the emission reported in
  Table~\ref{tab:cont_properties}. Contours are in 10\% steps of the
  peak value with the maximum contour at 90\%. The primary and
  synthesised beam sizes are shown as a circle and filled ellipse,
  respectively. The linear scale is illustrated by the line in the
  lower left of each image. The positions of the methanol maser spots
  are shown as crosses, with the same angular size in each image.}
\label{fig:g24.79_cont_maps}
\end{figure*}

\clearpage

\begin{figure*}
\begin{center}
\begin{tabular}{cc}
 \includegraphics[width=7.5cm, angle=-90, trim=0 0 -5 0]{figs/postscripts/g24.85/g24.85.1_1.mom0_2.ps} &
 \includegraphics[width=7.5cm, angle=-90, trim=0 0 -5 0]{figs/postscripts/g24.85/g24.85.18768.cont_h168.ps} \\
 \includegraphics[width=7.5cm, angle=-90, trim=0 0 -5 0]{figs/postscripts/g24.85/g24.85.34586.cont_h75.ps} &
 \includegraphics[width=7.5cm, angle=-90, trim=0 0 -5 0]{figs/postscripts/g24.85/g24.85.34592.cont_h168.ps} \\
\end{tabular}
\end{center}
\caption{G24.85+0.09 maps. The top-left image shows the integrated
  $\nhone$ emission as dashed contours and the 24GHz continuum
  emission as solid contours (taken from L07).  The numbers show the
  labels given to each of the cores. The remaining images show the
  continuum emission at each frequency and array configuration. The
  peak flux shown at the top of each image corresponds to the maximum
  value in that image rather than the fits to the emission reported in
  Table~\ref{tab:cont_properties}. Contours are in 10\% steps (except
  the bottom three contours in the bottom right image which are
  separated by 5\%) of the peak value with the maximum contour at
  90\%. The primary and synthesised beam sizes are shown as a circle
  and filled ellipse, respectively. The linear scale is illustrated by
  the line in the lower left of each image. The position of the
  methanol maser emission is shown as crosses, with the same angular
  size in each image.}
\label{fig:g24.85_cont_maps}
\end{figure*}

\clearpage

\end{document}